\documentstyle[aps]{revtex}
\newcommand{\f}{\begin{equation}}
\newcommand{\ff}{\end{equation}}
\begin{document}
\title{{\Large {\bf Physics with nonperturbative quantum gravity: \\  
radiation from a quantum black hole}}} 
\vskip.5cm
\author{
   Marcelo Barreira${}^1$\thanks{e-mail: marcelo@phyast.pitt.edu }, 
   Mauro Carfora${}^2$\thanks{e-mail: carfora@pavia.infn.it}, 
   Carlo Rovelli${}^1$\thanks{e-mail:  rovelli@pitt.edu}\\  
\vskip.2cm
  {\it  ${}^1$ Department of Physics and Astronomy, University of 
Pittsburgh,\\ 
 Pittsburgh PA 15260, USA\\  \vskip.2cm
${}^2$ Dipartimento di Fisica, Universita' di Pavia, Pavia, Italy}}
\date{\today}
\maketitle
\vskip.4cm
\begin{abstract}


\noindent We study quantum gravitational effects on black hole 
radiation, 
using loop quantum gravity.  Bekenstein and Mukhanov have recently 
considered the modifications caused by quantum gravity on Hawking's 
thermal black-hole radiation.  Using a simple ansatz for the eigenstates 
the area, they have obtained the intriguing result that the quantum 
properties of geometry affect the radiation considerably, yielding a 
definitely non-thermal spectrum.  Here, we replace the simple ansatz 
employed by Bekenstein and Mukhanov with the actual eigenstates of the 
area, computed using the loop representation of quantum gravity. We 
derive the emission spectra, using a classic result in number theory by 
Hardy and Ramanujan.  Disappointingly, we do not recover the 
Bekenstein-Mukhanov spectrum, but --effectively-- a Hawking's thermal 
spectrum.  The Bekenstein-Mukhanov result is therefore likely to be an 
artefact of the naive ansatz, rather than  a robust result. 
The result is an example of concrete (although 
somewhat disappointing) application of nonperturbative quantum gravity. 
\end{abstract}

\vskip1.3cm

\noindent
Quantum gravity research has traditionally suffered for a great scarcity 
of physical applications where theories and ideas could be tested, at
least in principle \cite{isham}.  
One of the few areas in which ideas on quantum gravity may be tested is 
black hole physics \cite{wald}.  The loop approach to quantum gravity 
\cite{carlolee} is now sufficiently developed 
that we may begin to probe it within ``physical'' applications.  It is 
thus natural to investigate what loop quantum gravity asserts about black 
hole physics. 

Recently, Bekenstein and Mukhanov \cite{bm} have suggested that the 
thermal nature of Hawking's radiation may be affected by quantum 
properties of gravity (For a review of earlier suggestions in this 
direction, see \cite{lee}).   As it is well known, Hawking derived the 
black hole thermal emission spectrum from quantum field theory in curved 
spacetime, therefore within the approximation in which the quantum 
properties of gravity are neglected.  Attempts have been made to relate 
Hawking's temperature with gravitational dynamics, but the problem of how 
quantum gravity affects black hole emission can be convincingly 
addressed only within a full theory of the quantum gravitational field.  
Bekenstein and Mukhanov observe that in most approaches  to quantum 
gravity the area can take only quantized  values \cite{garay}.  Since the 
area of the 
black hole surface is connected to the black hole mass, black hole mass 
is likely to be quantized as well.  The mass of the black hole decreases 
when radiation is emitted.  Therefore emission happens when the black 
hole makes a quantum leap from one quantized value of the mass (energy)
to a lower quantized 
value, very much as atoms do.  A consequence of this picture is that 
radiation is emitted at quantized frequencies, corresponding to the 
differences between energy levels.   Thus, quantum gravity implies 
a discretized emission spectrum for the black hole radiation.  

By itself, this result is not physically in contradiction with Hawkings 
prediction 
of a continuous thermal spectrum. To understand this, consider the black 
body radiation of a gas in a cavity, at high temperature.  This radiation 
has a thermal Planckian emission spectrum, essentially 
continuous.  However, radiation is emitted by elementary quantum emission 
processes yielding a discrete spectrum.  The solution of the apparent 
contradiction is that the spectral lines are so dense in the range of 
frequencies of interest, that they give rise --effectively-- to a 
continuous spectrum.  Does the same happen for a black hole?  

In order to answer this question, we need to know the energy spectrum of 
the black hole, which is to say, the spectrum of the Area. Bekenstein and 
Mukhanov pick up a simple ansatz: they assume that the Area is quantized 
in multiple integers of an elementary area $A_0$.  Namely, that the area 
can take the values
\f
                        A_n=n A_0,
\label{ansatz}
\ff
where $n$ is a positive integer, and $A_0$ is an elementary area of the 
order of the Planck Area 
\f
                A_0=\alpha\hbar G,
\ff
where $\alpha$ is a number of the order of unity ($G$ is Newton's 
constant and $c=1$).  Ansatz (\ref{ansatz}) is reasonable; it agrees, 
for instance, with the partial results on eigenvalues of the area in the 
loop representation given in \cite{weave}, and with the idea of a quantum 
picture of a geometry made by elementary ``quanta of area''.   Since the 
black hole mass is related to the area by
\f
                A = 16\pi G^2 M^2,
\ff
it follows from this relation and the ansatz (\ref{ansatz}) that the 
energy spectrum of the black hole is given by 
\f
        M_n = \sqrt{{n \alpha \hbar \over 16\pi G }}.
\label{mass}
\ff
Consider an emission process in which the emitted energy is much smaller 
than the mass $M$ of the black hole. From (\ref{mass}), the spacing 
between 
the energy levels is
\f
        \Delta M =  {\alpha\hbar  \over 32\pi G M}.
\ff
From the quantum mechanical relation $E=\hbar\omega$ we conclude that 
energy is emitted in frequencies that are integer multiple of the 
fundamental emission frequency 
\f
                \bar\omega={\alpha \over 32\pi G M}. 
\ff
This is the fundamental emission frequency of Bekenstein and Mukhanov 
\cite{bm} (they assume $\alpha=4\ln 2$).  Bekenstein and Mukhanov proceed 
in  \cite{bm} by showing that the emission amplitude remains the same as 
the one in Hawking's thermal spectrum, so that the full emission spectrum 
is given by spectral lines at frequencies multiple of $\bar\omega$,  
whose envelope is Hawking's thermal spectrum. 

As emphasized by Smolin in \cite{lee}, however, the Bekenstein-Mukhanov 
spectrum  is drastically different than the Hawking spectrum. Indeed, 
Hawking temperature is
\f
        T_H = {\hbar\over 8 \pi k GM}
\ff
($k$ is Botzmann constant); therefore the maximum of the Planckian 
emission 
spectrum of Hawking's thermal radiation is at
\f
        \omega_H \sim {2.82 k T_H \over\hbar} = {2.82\over 8\pi GM} 
        = {2.82\cdot 4\over\alpha}\, \bar\omega \approx \bar\omega.
\ff
That is: the fundamental emission frequency $\bar\omega$ is of the same 
order as  the maximum of the Planck distribution of the emitted 
radiation. It follows 
that there are only a few spectral lines in the regions where emission is 
appreciable. Therefore the Bekenstein-Mukhanov spectrum  is drastically 
different than the Hawking spectrum: the two have the same envelope, but 
while Hawking spectrum is continuous, the Bekenstein-Mukhanov spectrum is 
formed by just a few lines in the interval of frequencies where emission 
is appreciable. 

This result is of great interest because, in spite of its weakness, 
black hole radiation is still much closer to the possibility of
(indirect) investigation 
than any quantum gravitational effect of which we can think. Thus, a 
clear quantum gravitational signature on the Hawking spectrum is a 
very interesting effect. Is this Bekenstein-Mukhanov effect credible?

\vskip.3mm

One of the most definite results of loop quantum gravity is a calculation 
of the spectrum of the area from first principles \cite{area}.  Thus, 
following a suggestion in \cite{lee}, we may use loop quantum gravity to 
check the Bekenstein-Mukhanov result, by replacing the naive ansatz 
(\ref{ansatz}) with the precise spectrum computed in this approach to 
quantum gravity. 

Consider a surface $\Sigma$ --in the present case, the event horizon of 
the black hole--. According to loop quantum gravity, the area of $\Sigma$ 
can take only a set of quantized values. These quantized values are 
labelled by unordered n-tuplets of positive integers $\vec p = (p_1, ... 
, p_n)$ of arbitrary length $n$.  The spectrum is then given by
\f
        A_{\vec p} =  16\pi\hbar G
\sum_{i=1,n} \sqrt{{p_i\over2}\left({p_i\over2}+1\right)}, 
\label{spectrum}
\ff
For a full derivation of this spectrum, see 
\cite{area}.  The spectrum (\ref{spectrum}) is not complete. There is an 
additional sector corresponding to a class of ``degenerate'' states, 
whose physical interpretation is not obvious to us.  These degenerate 
states play no role in the present discussion, however.

If we disregard for a moment the term $+1$ under the square root in 
(\ref{spectrum}), we obtain immediately the ansatz (\ref{ansatz}), and 
thus the Bekenstein-Mukhanov result.  However, the $+1$ is there.  Let us 
study the consequences of its presence.  
First, let us estimate the number of Area eigenvalues between the value 
$A>>>l_0$ and the value $A+dA$ of the Area, where we take $dA$ much 
smaller than $A$ but still much larger than $l_0$.  Since the $+1$ in 
(\ref{spectrum}) affects in a considerable way only the tems with low 
$p_i$, we can neglect it for a rough estimate. Thus, we must estimate the 
number of unordered strings of integers $\vec p = (p_1, ... , p_n)$ such 
that \f
        \sum_{i=1,n} p_i = {A\over 8\pi\hbar G}>>1.
\ff
This is a well known problem in number theory.  It is called the 
partition problem. It is the problem of computing the number $N$ of ways 
in which an integer $I$ can be written as a sum of other integers.  The 
solution for large $I$ is a classic result by Hardy and Ramanujan 
\cite{Andrews}. According to the Hardy-Ramanujan formula, $N$ grows as the 
exponent of the square root of $I$. More precisely, we have for
large $I$ that   
\f
        N(I) \sim {1\over 4\sqrt{3}I} e^{\pi\sqrt{{2\over 3}I}}.
\ff
Applying this result in our case we have that the number of eigenvalues 
between $A$ and $A+dA$ is
\f
        \rho(A) \approx e^{\sqrt{\pi A\over 12\hbar G}}. 
\ff
Now, because of the presence of the $+1$ term, eigenvalues will overlap 
only accidentally: generically all eigenvalues will be distinct.  
Therefore, the average spacing between eigenvalues decreases 
exponentially with the inverse of the square of the area. This result is 
to be contrasted with the fact that this spacing is constant and of 
the order of the Planck area in the case of the naive ansatz 
(\ref{ansatz}). This conclusion is devastating for the 
Bekenstein-Mukhanov argument.  Indeed, the density of the energy levels 
becomes
\f
        \rho(M) \approx e^{\sqrt{4\pi G\over 3\hbar}M},
\label{density}
\ff
and therefore the spacing of the energy levels decreases {\it exponentially\ }  
with $M$.  It follows that for a macroscopical black hole the spacing 
between energy 
levels is infinitesimal, and thus the spectral lines are virtually dense 
in frequency. We effectively recover in this way Hawking's thermal 
spectrum (except, of course, in the case of a Planck scale black hole).
The conclusion is that the Bekenstein-Mukhanov effect disappears if we 
replace the naive ansatz (\ref{ansatz}) with the spectrum 
(\ref{spectrum}) computed from loop quantum gravity.  More generally, we 
have shown that the  Bekenstein-Mukhanov effect is strongly dependent on 
the peculiar form of the naive ansatz (\ref{ansatz}), and it is not 
robust. 
In a sense, 
this is a pity, because we loose a possible window on quantum geometry.

We have shown that the discretization of the spectrum derived by 
Bekenstein and Mukhanov disappears if we use quantitative result from loop 
quantum gravity. Our result 
indicates that loop quantum gravity is perhaps sufficiently 
mature to begin addressing concrete physical problems.

\end{document}